\documentclass[fleqn,usenatbib]{mnras}

\usepackage[T1]{fontenc}
\usepackage{ae,aecompl}

\usepackage[utf8]{inputenc}
\usepackage[english]{babel}
\usepackage{amsfonts}
\usepackage{animate}
\usepackage{lmodern}
\usepackage{rotating}
\usepackage{tikz}
\usepackage{mathtools}
\usepackage{hyperref}
\usepackage{url}
\usepackage{color}
\usepackage{journals}
\usepackage{graphicx}
\usepackage{amsmath}
\usepackage{amssymb}
\usepackage{xcolor}
\hypersetup{
    colorlinks,
    linkcolor={red!50!black},
    citecolor={blue!50!black},
    urlcolor={blue!80!black}
}

\author[M. Maturi et al.]{Matteo Maturi\thanks{maturi@uni-heidelberg.de},$^{1}$
  Fabio Bellagamba,$^{2,3}$
  Mario Radovich,$^{4}$
  Mauro Roncarelli,$^{2,3}$
  \newauthor
  Mauro Sereno,$^{2,3}$
  Lauro Moscardini,$^{2,3,6}$
  Sandro Bardelli,$^{2}$
  and Emanuella Puddu$^{5}$
  \\  $^{1}$ Zentrum für Astronomie, Universität Heidelberg, Philosophenweg 12, D-69120 Heidelberg, Germany\\
  $^{2}$ Dipartimento di Fisica e Astronomia, Alma Mater Studiorum -- Universit\`a di Bologna, via Piero Gobetti 93/2, I-40129 Bologna, Italy\\
  $^{3}$ INAF -  Osservatorio di Astrofisica e Scienza dello Spazio di Bologna, via Gobetti 93/3, I-40129 Bologna, Italy\\
  $^{4}$ INAF - Osservatorio Astronomico di Padova, vicolo dell’Osservatorio 5, I-35122 Padova, Italy\\
  $^{5}$ INAF - Osservatorio Astronomico di Capodimonte, Salita Moiariello 16, I-80131 Napoli, Italy\\
  $^{6}$ INFN - Sezione di Bologna, Viale Berti Pichat 6/2, I-40127 Bologna, Italy
}

\title{AMICO galaxy clusters in KiDS-DR3: sample properties and
  selection function}

\date{Submitted 2018}

\pubyear{2018}

\begin{document}

\maketitle

\begin{abstract}
  We present the first catalogue of galaxy cluster candidates derived
  from the third data release of the Kilo Degree Survey
  (KiDS-DR3). The sample of clusters has been produced using the
  Adaptive Matched Identifier of Clustered Objects (AMICO)
  algorithm. In this analysis AMICO takes advantage of the luminosity
  and spatial distribution of galaxies only, not considering
  colours. In this way, we prevent any selection effect related to the
  presence or absence of the red-sequence in the clusters. The
  catalogue contains $7988$ candidate galaxy clusters in the redshift
  range $0.1<z<0.8$ down to $S/N>3.5$ with a purity approaching $95$\%
  over the entire redshift range. In addition to the catalogue of
  galaxy clusters we also provide a catalogue of galaxies with their
  probabilistic association to galaxy clusters. We quantify the sample
  purity, completeness and the uncertainties of the detection
  properties, such as richness, redshift, and position, by means of
  mock galaxy catalogues derived directly from the data. This
  preserves their statistical properties including photo-$z$
  uncertainties, unknown absorption across the survey, missing data,
  spatial correlation of galaxies and galaxy clusters. Being based on
  the real data, such mock catalogues do not have to rely on the
  assumptions on which numerical simulations and semi-analytic models
  are based on. This paper is the first of a series of papers in which
  we discuss the details and physical properties of the sample
  presented in this work.
\end{abstract}


\section{Introduction}

clusters of galaxies are one of the fundamental probes to study the
nature of dark matter and dark energy \citep{umetsu15,
  planckSzCosmo16, Sartoris16, dehaan16, wang16, smith16, giocoli18,
  Schellenberger17, Corasaniti18}, gravity itself
\citep{Llinares12,LHuillier17}, to constrain neutrino masses
\citep{Costanzi13,Roncarelli15} as well as the far universe and the
early stages of star and galaxy formation when used as gravitational
lensing telescopes \citep{Zheng12, coe13, Bradley14, Kelly15,
  Rydberg18}. There are many ways to identify galaxy clusters: through
the X-ray emission
\citep{bohringer04,2008A&A...483..389P,piffaretti11,Merloni12,Clerc14},
the comptonization of the Cosmic Microwave Background photons by
the hot plasma they contain \citep{Reichardt13,Arnaud16,Hilton18}, the
gravitational lensing distortion they induce on background galaxies
\citep{2005A&A...442..851M,2007A&A...471..731P,2011MNRAS.413.1145B}
and the optical emission of their population of galaxies. Various
methods have been proposed and used for their detection in photometric
catalogue of galaxies. For instance wavelength filters
\citep{Gonzalez14,Benoist14}, friend-of-friends \citep{Farrens11},
methods based on Voronoi tessellation \citep{iovino14}, minimal
spanning trees \citep{Adami99}, red-sequenced finders
\citep{Rykoff14,Licitra16} and matched optimal filters \citep[][Adam
  et al. in prep.]{2011MNRAS.413.1145B,Bellagamba18}.

In this work, we searched for galaxy clusters in the third data
release of the Kilo Degrees Survey \citep[KiDS-DR3,][]{deJong17}.
With respect to our previous study on the second data release
\citep{Radovich16,Radovich17}, the analysis presented here benefits of
a larger survey area, better data quality, and significant
improvements in the cluster detection algorithm. For this task we use
the Adaptive Matched Identifier of Clustered Objects algorithm
\citep[AMICO,][]{Bellagamba18}, an optimal matched filter which takes
advantage of the known statistical properties of the field galaxies
and of galaxy clusters. Even if AMICO can deal with an arbitrary
number of quantities describing galaxies, in this specific application
we consider their spatial coordinates, magnitude, and photo-$z$
only. We deliberately avoided the use of their colors to aim at a
selection function minimally sensitive to the presence (or absence) of
the red-sequence of clusters.

To derive the uncertainty on the properties of the detections, the
purity and the completeness of the sample we realized a series of
realistic mock catalogues of galaxies based on the real KiDS data. In
doing so, we took care to preserve the actual masked areas in the
data, all photometric and photo-$z$ properties of the galaxies, as
well as their large scale correlation and the correlation of clusters
among themselves and inside large scale structures.

The structure of the paper is as follows. In Section~(\ref{sec:data})
we describe the data set. In Section~(\ref{sec:algorithm}) we
summarize the characteristics of the detection algorithm and the new
features used specifically for this work. The catalogue of galaxy
clusters and the comparison with existing catalogues are presented
in Sections~(\ref{sec:catalogs}) and (\ref{sec:other-catalogs}),
respectively. The uncertainties on the detection properties, the
completeness and purity of the sample are quantified in
Section~(\ref{sec:accessing}). Finally the conclusions are summarized
in Section~(\ref{sec:conclusions}).

\section{The data sets}\label{sec:data}

We analysed the galaxy catalogue coming from the KiDS Data Release 3
\citep{deJong17} obtained with the OmegaCAM wide-field imager
\citep{Kuijken11} mounted at the VLT Survey Telescope, a 2.6m
telescope sited at the Paranal Observatory
\citep[VST;][]{Capaccioli11}. OmegaCAM contains a mosaic of 32 science
CCDs offering a field of view of 1 deg$^2$ with a resolution of 0.21
arcsec/pixel. The data cover an area of about 440 $deg^2$ split into
two main stripes, one equatorial (KiDS-N) and one centred around the
South Galactic Pole (KiDS-S). The galaxy catalogue provides the
coordinates, the 2 arcsec aperture photometry in four bands (u, g, r,
i) and photometric redshifts for all galaxies down to the $5\sigma$
limiting magnitudes of 24.3, 25.1, 24.9 and 23.8 in the four bands,
respectively. We selected all galaxies with magnitude $r < 24$ for a
total of about $32$ million objects.

The photometric redshifts of the galaxies have been obtained with BPZ,
a Bayesian photo-$z$ estimator based on a template-fitting method
\citep{Benitez00,deJong17}. BPZ returns a photo-$z$ posterior
probability distribution function which is fully exploited by AMICO
(see below). Other two sets of photometric redshifts obtained via
Machine-Learning techniques, MLPQNA and ANNz2, are available in
KiDS-DR3 \citep{deJong17,Bilicki18}. An extensive analysis of the
probability distribution functions derived for these two sets will be
presented in \cite{Amaro18}: since it was still in progress when our
cluster catalog was derived, we opted to use the well tested BPZ
photometric redshifts. In next releases we will also investigate the
usage of Machine-Learning photometric redshifts.

\section{AMICO: the detection algorithm}\label{sec:algorithm}

For the detection of the galaxy clusters we used the AMICO code
\citep{Bellagamba18}. In this section we briefly summarize its main
concepts and the features recently implemented and adopted in the
following analysis.

\subsection{Linear optimal matched filtering}\label{sec:matchedfilter}

AMICO is based on a linear optimal matched filter approach
\citep{2005A&A...442..851M,2007A&A...462..473M,2010MNRAS.403..859V,2011MNRAS.413.1145B}. Within
this framework, the data, $d(\vec{x})=s(\vec{x})+n(\vec{x})$, are
modelled as the superimposition of the signal we are interested in,
i.e. the galaxy clusters signal $s(\vec{x})=A c(\vec{x})$, and a noise
component, $n(\vec{x})$, describing the contamination given by the
field galaxies. The filter itself is a kernel used to convolve the
data and it is derived through a constrained minimization procedure
aiming at estimating the signal amplitude, $A$, which is unbiased and
with minimum variance. Despite the fact AMICO can deal with an
arbitrary number of galaxy properties we restrict ourself to the
simple case in which the data points,
$\vec{x}_i=\left(\vec{\theta}_i,m_i,p_i(\vec{z})\right)$, are
individual galaxies, labelled with $i$, characterized by sky
coordinates, $\vec{\theta}_i$, an r-band magnitude, $m_i$, and a
photometric redshift distribution, $p_i(z)$. The aforementioned
convolution returning an estimate for $A$ is evaluated on a
three-dimensional grid ($\vec{\theta}_c$,$z_c$), with resolution of
$0.3\arcmin$ across the sky and $0.01$ in redshift, and is discretized
to deal with counts of galaxies:
\begin{equation}
  \label{eq:amplitude}
  A(\vec{\theta}_c,z_c) = \alpha^{-1}(z_c) \sum_{i=1}^{N_{gal}}\frac{C(z_c;\vec{\theta}_i-\vec{\theta}_c,m_i)p_i(z_c)}{N(m_i,z_c)} - B(z_c) \;.
\end{equation}
Here, $N$ and $C$ describe the properties of the field and cluster
galaxies at redshift $z_c$, respectively, as it will be detailed in
Section~(\ref{sec:model}); the factor $\alpha$ takes care of the
filter normalization and $B$ is a background subtraction term
quantifying the average contribution of the field galaxies to the
total signal amplitude. The expected r.m.s. of the amplitude is given
by
\begin{equation}
  \label{eq:amplitudesigma}
  \sigma_A(\vec{\theta}_c,z_c) = \alpha(z_c)^{-1} + A(\vec{\theta}_c,z_c)\frac{\gamma(z_c)}{\alpha(z_c)^{2}}\;,
\end{equation}
where the first term refers to the stochastic fluctuations of the
background and the second one is related to the Poissonian
fluctuations given by the galaxies of a cluster with amplitude
$A$. The factors $B$, $\alpha$, and $\gamma$ are properties of the
filter and solely depend on the cluster\footnote{We base the redshift
  distribution of the model on the average $P(z)$ of the input
  galaxies as detailed in \cite{Bellagamba18}.} and field models. The
definition of $B$ is given in \cite{Bellagamba18}, while the new
definitions of $\alpha$ and $\gamma$, implementing the new features of
the algorithm, are given in Section~(\ref{sec:new}).

\begin{figure}
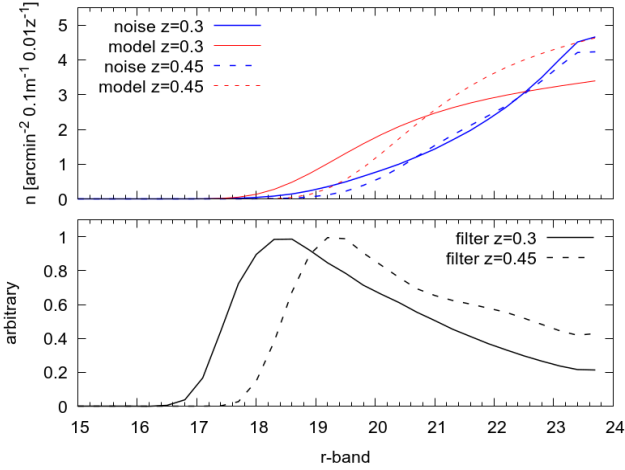

  \centering
  \includegraphics[width=0.48\textwidth]{{{./fig/filter_lum_z0.3-0.45}}}
  \caption{Top panel: the luminosity function of the models of the
    field galaxies in blue (the noise component) and the one of the
    cluster members in red (the cluster model) at redshift $z=0.35$
    (solid lines) and at redshift $z=0.45$ (dashed lines). Bottom
    panel: the magnitude distribution of the resulting filters for the
    same two redshifts.  }
  \label{fig:model}
\end{figure}

\begin{figure*}
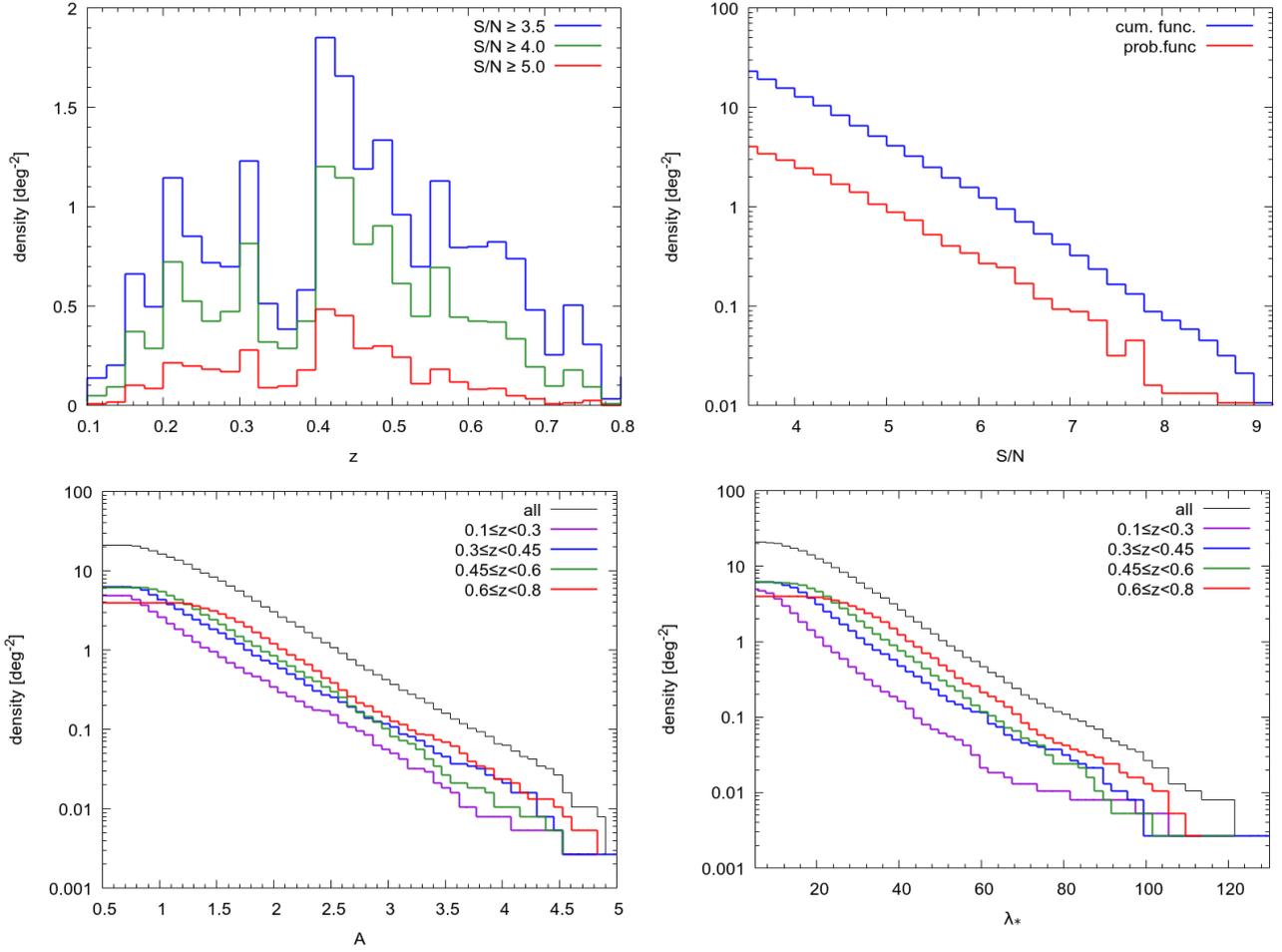

  \centering
  \includegraphics[width=0.48\textwidth]{{{./fig/all_z_distribution_data}}}
  \includegraphics[width=0.48\textwidth]{{{./fig/all_z_SN_data}}}
  \includegraphics[width=0.48\textwidth]{{{./fig/all_z_amplitude_data}}}
  \includegraphics[width=0.48\textwidth]{{{./fig/all_z_richstar_data}}}
  \caption{The cumulative distributions of the number density of
    detections as a function various properties. In the top left
    panel, as a function of redshift and for three different
    signal-to-noise ratios; in the top right panel, the distribution
    as a function of signal-to-noise ratio (red curve) and its
    cumulative (blue curve); in the bottom panels, as a function of
    amplitude $A$ (left) and intrinsic richness $\lambda_*$ (right).}
  \label{fig:cat_distributions}
\end{figure*}

Once the amplitude $A$ has been evaluated for all angular and
redshift positions, the first cluster candidate is then identified as
the location with the largest likelihood,
\begin{equation}
  \label{eq:likelihood}
  \mathcal{L}(\vec{\theta}_c,z_c) = \mathcal{L}_0 + A^2(\vec{\theta}_c,z_c)\alpha(z_c)\;,
  \Delta\mathcal{L}(\vec{\theta}_c,z_c) = A^2(\vec{\theta}_c,z_c)\alpha(z_c)\;.
\end{equation}
and positive amplitude. Here, $\mathcal{L}_0$ is a constant of no
relevance that we no further discuss. With a cluster detection at
hand, labelled with $j$, we evaluate the probability of all galaxies
in that region of being its members,
\begin{equation}
  \label{eq:probability}
  P_i(j) = P_{f,i} \frac{A_j C_j(\vec{\theta}_i-\vec{\theta}_j,\vec{m}_i)p_i(z_j)}
  {A_j C(\vec{\theta}_i-\vec{\theta}_j,\vec{m}_i) p_i(z_j) + N(\vec{m}_i,z_j)}  \;.
\end{equation}
Here, 
\begin{equation}
  \label{eq:prob_field}
  P_{f,i} \equiv 1-\sum_j P_i(j) \;,
\end{equation}
is the probability of the $i$-th galaxy to belong to the field. In
general, a galaxy can be associated to more detections because they
can overlap. We store the information of all cluster members down to a
membership probabilistic association of $P_i(j)=0.0$.

To proceed with the search of further clusters, we remove from the
amplitude map the contribution of the last found detection,
re-evaluate the likelihood and variance to finally identify a new
candidate as done with the previous one. The removal of a detection is
done by taking advantage of the membership probabilistic association
of the galaxies to a detection, $P_i(j)$, as follows
\begin{equation}
  \label{eq:amplitude_new}
  A_{new}(\vec{\theta}_j,z_k) = A(\vec{\theta}_j,z_k) -
  \sum_{i=1}^{N_{gal}} P_i(j) \frac{C_j(\vec{\theta}_i-\vec{\theta}_j,\vec{m}_i) p_i(z_k)}{ N(\vec{m}_i,z_k)}\;.
\end{equation}
This signal subtraction facilitates a better identification of objects
which might be blended with those with larger amplitudes. We refer to
this process as ``{\it cleaning}''. This iterative process proceeds
down to a desired minimum signal-to-noise ratio, $S/N\coloneqq
A/\sigma_A$, that in this work is set to $(S/N)_{min}=3.0$.

\begin{table}
  \centering
  \begin{tabular}{lll}
    name & description\\
    \hline
    $A$&amplitude, natural output of the filter\\
    $\lambda$&apparent richness, number of visible galaxies\\
    $\lambda^*$&intrinsic richness, as $\lambda$ but for $r<R_{200}$ and $m<m_*+1.5$\\
    \hline
  \end{tabular}
  \caption{The three mass proxy delivered by AMICO.}
  \label{tab:proxies}
\end{table}

\subsection{New features of the algorithm}\label{sec:new}

In order to correctly normalise the amplitude $A$ and estimate its
uncertainty $\sigma_A$, AMICO calculates the quantities $\alpha(z_c)$
and $\gamma(z_c)$, which depend on the properties of the redshift
probability distributions of the galaxy sample. In
\cite{Bellagamba18}, this was done by introducing $q(z_c, z)$, the
typical redshift probability distribution for a galaxy which lies at
redshift $z_c$, computed as
\begin{equation}\label{eq:qz}
q(z_c,z) = \left(\sum_{i=1}^{N_\text{gal}} p_i(z_c)  \right)^{-1}  \sum_{i=1}^{N_\text{gal}} p_i(z-z_c+z_\text{peak,i})\ p_i(z_c) ,
\end{equation}
where $z_{\text{peak},i}$ is the most probable redshift for the $i$-th
galaxy.  In this analysis, we refined this treatment in two
ways. First of all, we computed the photo-$z$ properties as a function
of the $r$-band magnitude to capture the different precision of
photo-$z$s depending on the quality of the galaxy photometry. Then, we
replaced $q$ with two different statistics $q_1$ and $q_2$ defined by
\begin{equation}
  q_1(m,z_p,z_c) = \left(\sum_{z_{\text{peak},i}=z_p} p_i(z_p)  \right)^{-1} \sum_{z_{\text{peak},i}=z_p} p_i(z_p)\ p_i(z_c) ,
\end{equation}
and
\begin{equation}
  q_2(m,z_c,z_p) = \left(\sum_{i=1}^{N_\text{gal}} p_i(z_c)  \right)^{-1} \sum_{z_{\text{peak},i}=z_p} p_i(z_c)\ p_i(z_p) ,
\end{equation}
where $z_{\text{peak},i}=z_p$ means that the sum runs only on the
galaxies whose peak corresponds to $z_p$.  In practice, $q_1$
describes the typical $p(z)$ that peaks at $z_p$, while $q_2$
describes the probability distribution for the peak, $z_p$, of a galaxy
that is located at redshift $z_c$. Together, they allow to measure the typical
precision of the redshift probability distribution as a function of $z$,
but also the small-scale features of the $p(z)$-peaks distribution,
removing the smoothing that is implicit in Equation~(\ref{eq:qz}).
With these two new quantities, the constants $\alpha(z_c)$ and
$\gamma(z_c)$ can be now defined as
\begin{equation}\label{eq:alpha}
  \alpha(z_c) = \int \frac {M_c^2(\vec \theta - \vec \theta_c, m)\ q_1(m,z_p,z_c) q_2(m,z_c,z_p)} {N(m,z_c)}\ d^2\theta\ dm\ dz_p
\end{equation}
and
\begin{equation}\label{eq:gamma}
  \gamma(z_c) = \int \frac {M_c^3(\vec \theta - \vec \theta_c, m)\ q^2_1(m,z_p,z_c) q_2(m,z_c,z_p)} {N^2(m,z_c)}\ d^2\theta\ dm\ dz_p.
\end{equation}

\subsection{Mass proxies and cluster richness}

As discussed in Section~(\ref{sec:matchedfilter}), the natural output
of the linear optimal matched filter is the amplitude, $A$, expressed
by Equation~(\ref{eq:amplitude_new}). In this Section, we derive two
other mass proxies based on the probabilistic membership association
of the galaxies to detections (Equation~\ref{eq:probability}). The
first one is the apparent richness that is defined as the sum of the
probabilities of all galaxies associated to the $j-th$ detection,
\begin{equation}
  \label{eq:apprichness}
  \lambda_{j} = \sum_{i=1}^{N_{gal}} P_i(j) \;.
\end{equation}
This quantity represents the number of visible galaxies belonging to a
detection. Clearly, this number depends on the cosmic distance at
which a cluster is located so that $\lambda$ is a redshift dependent
quantity. In fact, the further the cluster, the smaller the number of
visible members. The advantage of this definition with respect to the
amplitude $A$ is that it is related to a direct observable, namely the
number of visible galaxies.

The second mass proxy here used is the intrinsic richness defined in a
similar fashion but by summing over the galaxies brighter than
$m_*+1.5$ and within the virial radius, $R_{200}$,
\begin{equation}
  \label{eq:starrichness}
  \lambda_{*j} = \sum_{i=1}^{N_{gal}} P_i(j) \quad\mbox{with}\quad
  \left\{
  \begin{array}{lr}
    m_i<m_*(z_j)+1.5 \\
    r_i(j) < R_{200}(z_j)
  \end{array}
  \right.\;.
\end{equation}
The radial cut $R_{200}$ and $m_*$ are parameters of the model we used
for the construction of the filter, see Section~(\ref{sec:model}),
that we adopt for internal consistency. Obviously each detection has
its own $R_{200}$ and we could scale the radial cut-off with the
detection amplitude $A$ or apparent richness $\lambda$, but this would
lead to an additional stochastic scatter given by the uncertainty of
these quantities. Note that these definitions depend on the conversion
from angular to physical distances and therefore imply the assumption
of a specific cosmological model\footnote{We assume a $\Lambda$CDM
  cosmological model with $\Omega_\Lambda=0.7$, $\Omega_m=0.3$,
  $h=0.7$ through out the paper.}. Despite that, $\lambda_*$ is a
nearly redshift independent quantity which can better characterize the
clusters mass. This is because the threshold $m_*+1.5$ is well below
the magnitude limit for the galaxy sample for the entire redshift
range considered in this work. In Table~(\ref{tab:proxies}), we list
the mass proxies provided by AMICO and delivered with the catalogue of
galaxy clusters.

\subsection{Model description: cluster and field galaxies}\label{sec:model}

The cluster model, $C(z_c;r,m)$, describes the expected galaxy
distribution as a function of distance from the centre, $r =
|\vec{\theta}_i-\vec{\theta}_c|$, and $r$-band magnitude, $m$, for a
cluster at redshift $z_c$. In this work, the cluster model is
constructed from a luminosity function $\Phi(m)$ and a radial profile
$\Psi(r)$ as
\begin{equation}
  C(r,m) = \Phi(m)\Psi(r) \;,
\end{equation}
where we made implicit the dependence on the redshift $z_c$. The
parameters for these distributions are taken from the analysis of a
sample of Sunyaev Zel’dovich detected clusters observed by the Dark
Energy Survey \citep[DES;][]{Hennig17,Zenteno16}. These clusters cover
a redshift range $0 <z <1.1$, which is broadly comparable to ours, and
their detection via the Sunyaev Zel’dovich effect (SZ, hereafter)
avoids any selection bias related to the optical properties of the
galaxies in clusters, which could introduce systematics in the
detection process.

In particular, the luminosity function $\Phi(m)$ follows a Schechter
function \citep{schechter76}
\begin{equation}
  \Phi(m) = 10^{-0.4(m-m_\star)(\beta+1)} \exp{\left[-10^{-0.4(m-m_\star)}\right]} \;.
\end{equation}
Note that we only use the shape of the distribution and not the
normalization since the latter is absorbed by the constants during the
filter construction. The typical magnitude $m_\star$ as a function of
redshift is derived from a stellar population evolutionary model with
a decaying starburst at redshift $z$ = 3 (decay time = 0.4 Gyr) and a
Chabrier initial mass function \citep[IMF;][]{Bruzual03}. This model
has been described and confirmed in DES data by \cite{Zenteno16} where
they also derived a mean faint-end slope $\beta$ of $-1.06$, which we
adopt.

\begin{figure}
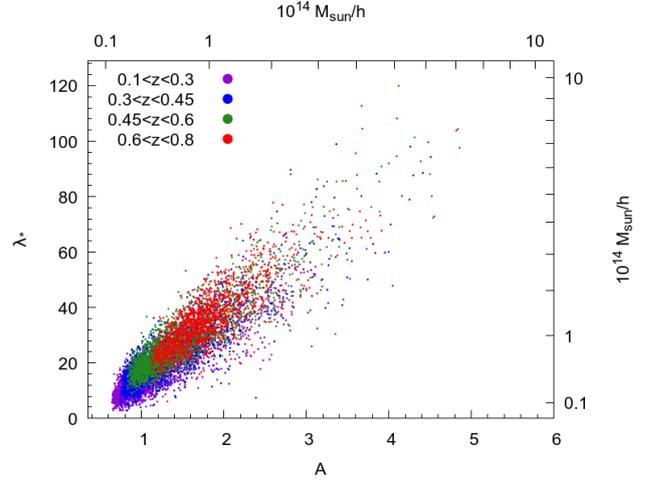

  \centering
  \includegraphics[width=0.48\textwidth]{{{./fig/all_amplitude_richstar_data_zrange}}}
  \caption{Correlation between the amplitude $A$ as returned by the
    matched filter and the intrinsic richness $\lambda_*$ defined as
    the sum of the membership probabililistic association of the
    galaxies (with $m<m_*+1.5$ and within $r<r_{200}$ from the
    detection) to clusters. Each cluster is color-coded according to
    its redshift, as labeled in the figure. The masses indicated in
    the secondary axis results from the scaling relations derived in
    Bellagamba et al. submitted.}
  \label{fig:amp-rich}
\end{figure}

For the radial profile $\Psi(r)$ we assume an Navarro-Frenk-White
profile \citep[NFW;][]{1997ApJ...490..493N}
\begin{equation}
\Psi(r) = \frac {C_0}{\displaystyle \frac {r}{r_s} \biggl(1+\frac{r}{r_s}\biggr)^2} \;,
\end{equation}
where the scale radius depends on the concentration $c$ via $r_s\equiv
R_{200}/c_{200}$. \cite{Hennig17} found the NFW distribution is a good
description of the observed cluster profiles, with a mean $c$ =
3.59. For our cluster model we used this value and an $R_{200}$
corresponding to a mass $M_{200}$ = $10^{14} M_\odot/h$, typical for
the cluster sample we want to target. The normalisation parameter
$C_0$ is such that the total number of galaxies $N_{200}$ inside
$R_{200}$ and below $m_\star+2$ is coherent with the relation with
$M_{200}$ found by \cite{Hennig17}. For a mass $M_{200}$ = $10^{14}
M_\odot/h$, this corresponds to $N_{200}$ = 22.9.

The field galaxies distribution $N(m,z_c)$ can be approximated by the
total distribution in the galaxy sample, as the cluster component is
small. For each redshift $z_c$, we build $N(m,z_c)$ weighing each
galaxy with its redshift probability distribution $p(z_c)$.

For illustration purposes, we show in the top panel of
Figure~(\ref{fig:model}) the luminosity function of both field
galaxies and cluster members at redshift $z=0.35$ (solid lines) and
$z=0.45$ (dashed lines). The magnitude dependence of the algorithm
filter resulting from the use of these luminosity functions is shown
in the bottom panel of the same figure. Such filter turns out to be a
band-pass filter with gives more weight to the galaxies with a certain
luminosity on the bright end side, the higher the redshift, the larger
the magnitude which the filter peaks at.

\begin{figure*}
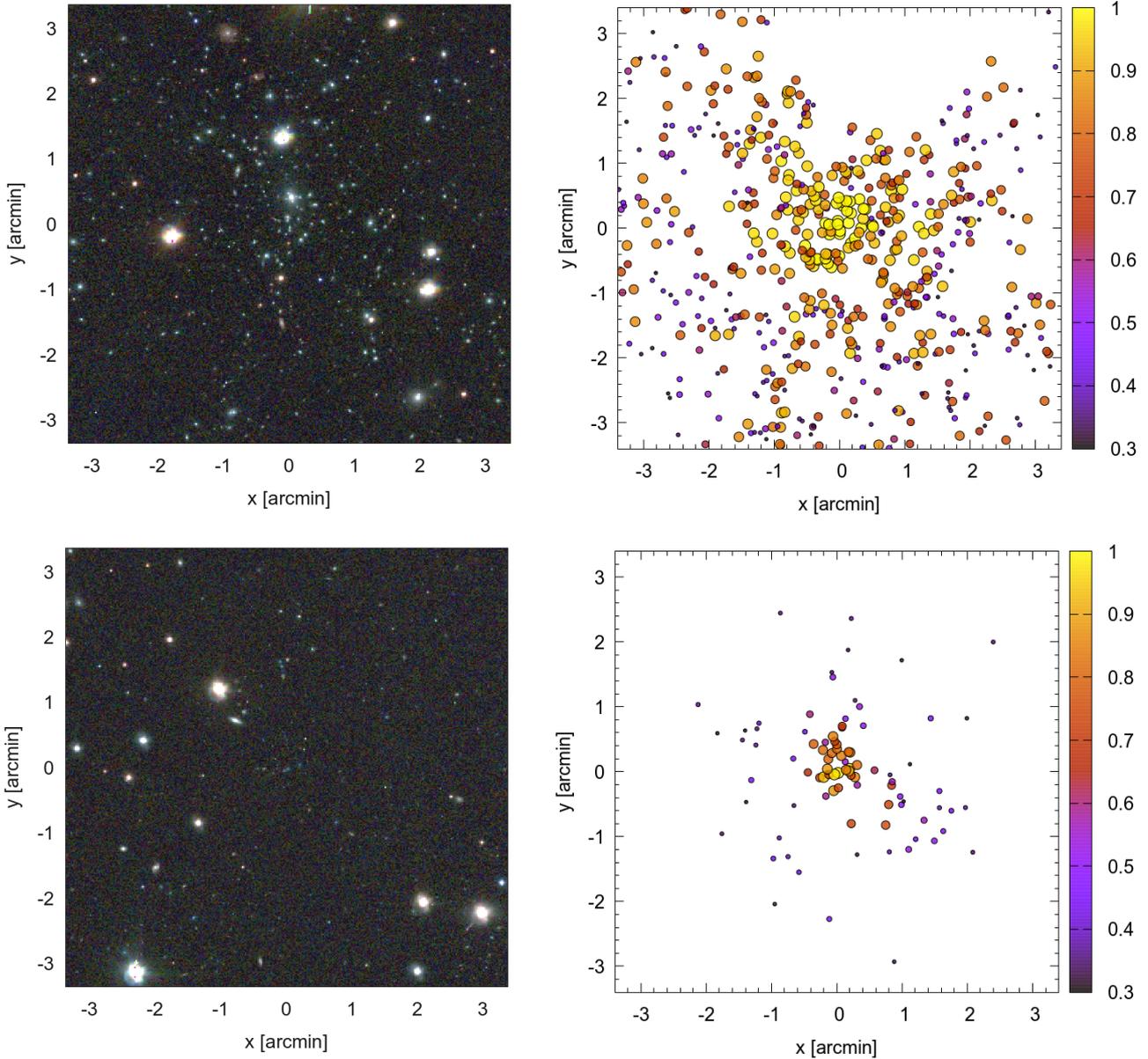

  \centering
  \includegraphics[width=0.46\textwidth]{{{./fig/stamp_4141_contrast}}}
  \includegraphics[width=0.53\textwidth]{{{./fig/KIDS_210.6_2.5_1_members}}}
  \
  \includegraphics[width=0.46\textwidth]{{{./fig/stamp_3934_contrast}}}
  \includegraphics[width=0.53\textwidth]{{{./fig/KIDS_187.5_2.5_13_members}}}
  \caption{In the left panels we show the colour composite (g,r,i)
    image of a rich cluster, with $\lambda_*=121$, at $z=0.28$,
    detected with $S/N=10.4$ (top panel) and a small one, with
    $\lambda_*=49.22$, at a higher redshift, $z=0.69$, detected with
    $S/N=4.6$ (bottom panel). Both stamps are centred at the position
    identified by AMICO and have a size of $400$ arcsec on a side. In
    the right corresponding panels we show the galaxies in these
    fields sized and coloured according to their probabilistic
    association to the detection (colour-coded as in the bar on the
    right).}
  \label{fig:members-circles}
\end{figure*}

\section{Clusters detectitons and galaxy members}\label{sec:catalogs}

In this section we describe the catalogue of galaxy clusters detected in
the KiDS-DR3 with the AMICO code. We provide the main statistical
properties of the sample and present few examples of detections.

\subsection{The catalog of galaxy clusters}

The survey covers a total area of $438$ deg$^2$ but we rejected all
galaxies falling in those regions severely affected by satellite
tracks, haloes produced by bright stars, and image artefacts in
general, leaving us with $414$ deg$^2$ \citep[``Primary halo'' masks,
  see][]{deJong15}. The remaining galaxies have been used to produce
an initial set of detections down to $S/N\geq 3.0$ as explained in
Section~(\ref{sec:algorithm}). This first sample of detections has
been then filtered out by rejecting all objects falling in the more
restrictive masks used for the weak lensing analysis
\citep[``Secondary/tertiary halo'' masks, see][]{deJong15}. The final
effective area is of $377$ deg$^2$, i.e. $86$\% of the total area of
the survey. All detections with $S/N>3.0$ have been used for the
construction of the mock simulations discussed in
Section~(\ref{sec:mocks}), but for the final catalogue of clusters we
kept only those with $S/N>3.5$ obtaining a final sample with $7988$
objects\footnote{The catalogue is available on request.}. This buffer
in $S/N$ is necessary for constructing reliable mock catalogues and
derive solid statistical properties for our cluster sample as it will
be detailed in Section~(\ref{sec:mock-clusters}). The entries of the
catalogue are specified in Table~(\ref{tab:entries_list}). For all
6972 objects falling in the redshift range $0.1<z<0.6$, we also
provide mass estimates obtained via stacked weak-gravitational lensing
(Bellagamba et al. submitted).

In Figure~(\ref{fig:cat_distributions}) we summarize the main
statistical properties of the detections listed in the catalogue by
showing their number density as a function of redshift (top left
panel), signal-to-noise ratio (top right panel), amplitude $A$ and
intrinsic richness $\lambda_*$ (left and right bottom panels,
respectively). The drop in density at $z\approx 0.38$ is due to
problems in the returned photometric redshift due to the shape of the
g and r filters, resulting in a not optimal covering of the $4000\AA$
break at that redshift \citep[see e.g.,][]{Padmanabhan05}.

Finally, Figure~(\ref{fig:amp-rich}) shows the amplitude $A$ against
the intrinsic richness $\lambda_*$ for the galaxy clusters falling in
four different redshift bins. The mass indicated on the secondary axes
is derived from the scaling relations of $\lambda_*$ and $A$ based on
weak-lensing measurements and based on a fiducial value of $z=0.35$
(Bellagamba et al. submitted). We can safely show the mass based on a
scaling relation computed at a specific redshift because of its small
redshift dependence.

\begin{figure}
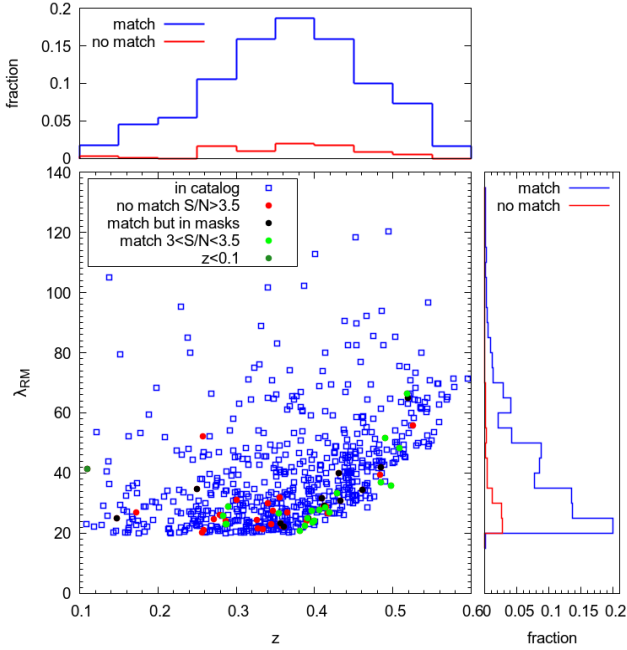

  \centering
  \includegraphics[width=0.5\textwidth]{{{./fig/rm_z_lambda_all}}}
  \caption{The correlation of richness $\lambda_{RM}$ and redshift $z$
    as measured with redMaPPer of matched (blue curves) and
    non-matched detections (red curves). Most of the non-matched
    detections are close to the detection limit of redMaPPer. Some of
    the detections are matched but they fall at the very border of
    masked areas (black circles), a bit below the minimum $S/N=3.5$ we
    considered (green circles) or because outside the redshift range
    (dark green circle).}
  \label{fig:RM_all}
\end{figure}

\begin{figure}
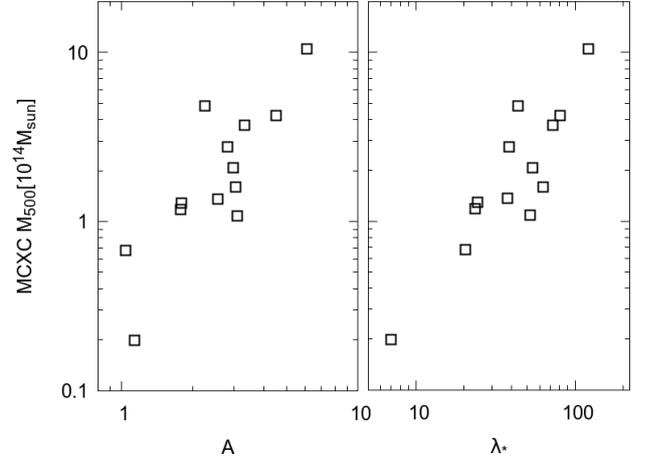

  \centering
  \includegraphics[width=0.48\textwidth]{{{./fig/MCXC}}}
  \caption{Correlation between the mass estimates reported in the MCXC
    catalogue against the amplitude $A$ (left panel) and against the
    intrinsic richness $\lambda_*$ (right panel).}
  \label{fig:MCXC_mass}
\end{figure}

\begin{figure}
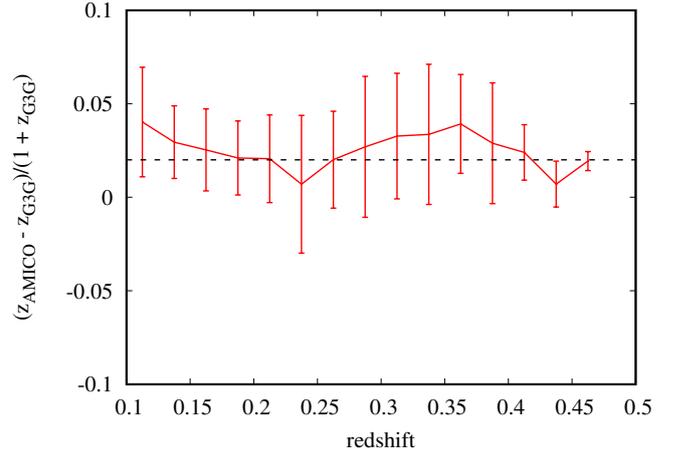

  \centering
  \includegraphics[width=0.5\textwidth]{{{./fig/G3G_match_red_bias}}}
  \caption{The difference between the clusters redshifts measured with
    AMICO and the spectroscopic ones taken from the GAMA G$^3$C
    catalogue. The bias affecting the KiDS photometric redshifts is
    evident.}
  \label{fig:gama_z}
\end{figure}

As an example of detections, we show in the left panels of
Figure~(\ref{fig:members-circles}) the image cut-outs of a rich
cluster, with $\lambda_*=121$, at relatively low redshift, $z=0.28$,
and an intermediate one, with $\lambda_*=49.22$, at a higher redshift,
$z= 0.69$. The first cluster has been detected with $S/N=10.4$ while
the second one with $S/N=4.6$. Both cut-outs are of $400$ arcsec in
size. In the corresponding right panels we show how the AMICO code
``sees'' the same two clusters. The circles mark the position of the
galaxies visible in the image cut-outs with size and color related to
their probabilistic association to the detection. The plot shows
clearly the depenence of membership probabilistic association with the
angular distance from the detection center. However, the value of
$P_i(j)$ has also a dependence on magnitude and on the $p(z)$ of the
galaxy (see eq. 4). Other examples are shown in
Appendix~(\ref{sec:stamps_examples}).

\subsection{The catalogue of cluster members}\label{sec:catOfMembers}

As discussed in Section~(\ref{sec:matchedfilter}), AMICO returns a
probabilistic association of galaxies to cluster candidates, see
Equatiton~(\ref{eq:probability}). Using this information, we present a
catalogue of cluster members with a maximum of twenty associations to
clusters, labelled with $j$, as well as their probability to be field
galaxies, $P_{f,i}$. We note that in the catalogue we do not apply any
cut in membership probability. This catalogue can be used to study the
properties of clusters, galaxy formation, help in the removal of the
foreground for weak-lensing studies, improve the strong lensing
estimates derived with photometric based algorithms
\citep{Stapelberg18, carrasco18}, among many other applications.

\section{Correlation with other data sets}\label{sec:other-catalogs}

In this section we compare our sample with other catalogues of galaxy
clusters published in the literature. A positive match between one of
our detections and one of the other entries occurs if they lay within
$\Delta z=0.1$ and $\Delta R=1$Mpc/h one from the other.

\subsection{RedMaPPer optically-selected clusters}

We compared our detections with the cluster candidates listed in
\cite{Rykoff14}. This catalogue has been obtained by running the
redMaPPer algorithm on the Sloan Digital Sky Survey Data Release 8
(SDSS-DR8) and contains entries within the redshift range
$0.08<z<0.55$. Of the 681 detections falling in the un-masked regions
of KiDS-DR3 data, 624, i.e. $92\%$, find a match. Of the remaining 57
detections, 20 have been detected by AMICO with a signal-to noise
ratio in the range $3<S/N<3.5$ below our restrictive threshold of
$S/N=3.5$, while other 13 satisfy the matching criteria but lay inside
a KiDS masked area. The remaining 24, i.e. less then $4\%$ of the
redMaPPer sample, do not have a counterpart with $S/N>3$. These
redMaPPer cluster candidates have a richness measured by redMaPPer,
$\lambda_{RM}$, close to the detection limit of this algorithm as
illustrated in Figure~(\ref{fig:RM_all}). Since the KiDS and the SDSS
data sets are very different in terms of depth and image quality, no
further comparison would be meaningful.

\begin{table}
  \caption{Clusters of galaxies listed in the Planck PSZ2 catalogue for
    which the redshift information was previously missing. The
    subscript `p' indicates the values listed in the Planck catalogue
    while `a' those listed in our catalogue.}  \centering
  \begin{tabular}{lllllll}
    ID$_{\mbox{p}}$ & RA$_{\mbox{p}}$ & DEC$_{\mbox{p}}$ & ID$_{\mbox{a}}$ & RA$_{\mbox{a}}$ & DEC$_{\mbox{a}}$ & z$_{\mbox{a}}$\\
    \hline
    11 & 358.351 & -33.2932 & 7485 & 358.385 & -33.2837 & 0.67\\
    19 & 350.538 & -34.5752 & 6948 & 350.468 & -34.6173 & 0.23\\
    39 & 354.054 & -32.1343 & 7226 & 354.053 & -32.1320 & 0.41\\
    41 & 342.976 & -33.3942 & 6473 & 342.963 & -33.4027 & 0.24\\
    44 & 356.853 & -31.1509 & 7361 & 356.884 & -31.1470 & 0.45\\
    48 & 341.633 & -32.2011 & 6295 & 341.717 & -32.2280 & 0.50\\
    50 & 351.169 & -30.6511 & 7002 & 351.156 & -30.6723 & 0.31\\
    59 & 340.637 & -30.3150 & 6241 & 340.618 & -30.4084 & 0.24\\
    1033 & 39.6692 & -30.8391 & 7601 & 39.6215 & -30.8968 & 0.55\\
    1606 & 216.108 & -2.73976 & 4120 & 216.089 & -2.83463 & 0.77\\
    \hline
  \end{tabular}
  \label{tab:planck_SZ}
\end{table}

\subsection{Planck SZ-selected clusters}\label{sec:planck}

Of all clusters listed in the second Planck catalogue of
Sunyaev-Zeldovich sources \citep{Arnaud16}, 19 fall in the area we
processed and all of them have been detected by AMICO. The redshift of
10 of these objects has not been reported in the literature to our
knowledge and for this we list in Table~(\ref{tab:planck_SZ}) our
estimate as measured with the AMICO code. Note that the cluster with
ID$_{\mbox{p}}=1606$ has not been detected with redMaPPer because
located at a redshift, $z=0.77$, which exceeds their maximum limit.

\subsection{MCXC X-ray selected clusters}\label{sec:MCXC}

We then compared our mass proxies with the X-rays mass estimates
listed in the Meta-Catalog of X-ray detected Clusters of galaxies
\citep[MCXC;][]{piffaretti11}. The MCXC catalogue comprises X-rays
selected clusters collected in archival data and includes the ROSAT
All Sky Survey-based (NORAS, REFLEX, BCS, SGP, NEP, MACS, and CIZA)
and serendipitous cluster catalogues (160SD, 400SD, SHARC, WARPS, and
EMSS) for a total of 1743 objects. Since the data have been taken with
different instruments and exposure times, they have been homogenized
in order to provide a coherent picture for this sample. All 13
clusters of the MCXC catalogue falling in the KiDS-DR3 foot-print have
been identified by AMICO. In Figure~(\ref{fig:MCXC_mass}) we compare
our two mass proxies, $A$ and $\lambda_*$, with the value of $M_{500}$
derived with the X-Rays observations. A well defined correlation is
evident.

\subsection{GAMA spectroscopy}\label{sec:gama}

We finally used the GAMA-I galaxy group catalogue
\citep[G$^3$C;][]{Driver09,Driver11,Liske15} to verify the redshift
estimate of the clusters provided by the AMICO code. GAMA is a highly
complete spectroscopic survey up to a Petrosian r-band magnitude of
$19.8$ and comprises $110.192$ galaxies, $40\%$ of which belong to
$14.388$ galaxy groups identified with a friends-of-friends (FoF)
algorithm in the redshift range $0\le z\le 0.5$ \citep{Robotham11}. In
Figure~(\ref{fig:gama_z}) we plot the relative scatter between the
redshift estimates of the groups as measured by AMICO and those listed
in the G$^3$C catalogue. A clear bias, well described by $\Delta
z/(1+z) \sim 0.02$, emerges from this comparison. This bias
corresponds to what was found by \cite{deJong17} when comparing KiDS
photo-zs with GAMA spec-z (See their table 8). Since the sample of the
G$^3$C is limited to $z<0.5$, we can not draw any conclusions for
clusters at higher redshifts. More details regarding this bias and how
we deal with it will be given in Section~(\ref{sec:uncertainties})
together with an extensive discussion of all other uncertainties.

\begin{figure}
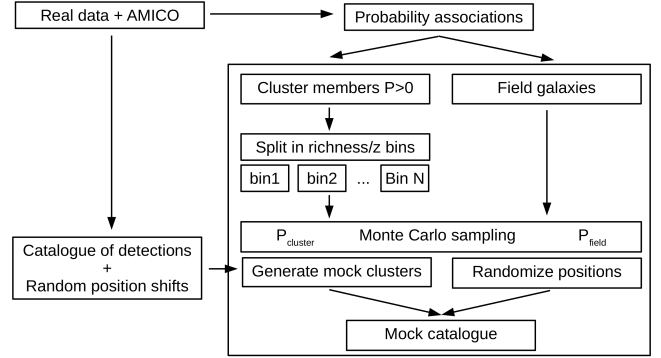

  \centering
  \includegraphics[width=0.48\textwidth]{{{./fig/scatch_mocks_black_cut_v2}}}
  \caption{Flow chart showing the process used to create the mock
    simulations.}
  \label{fig:scatch-mock}
\end{figure}

\begin{figure*}
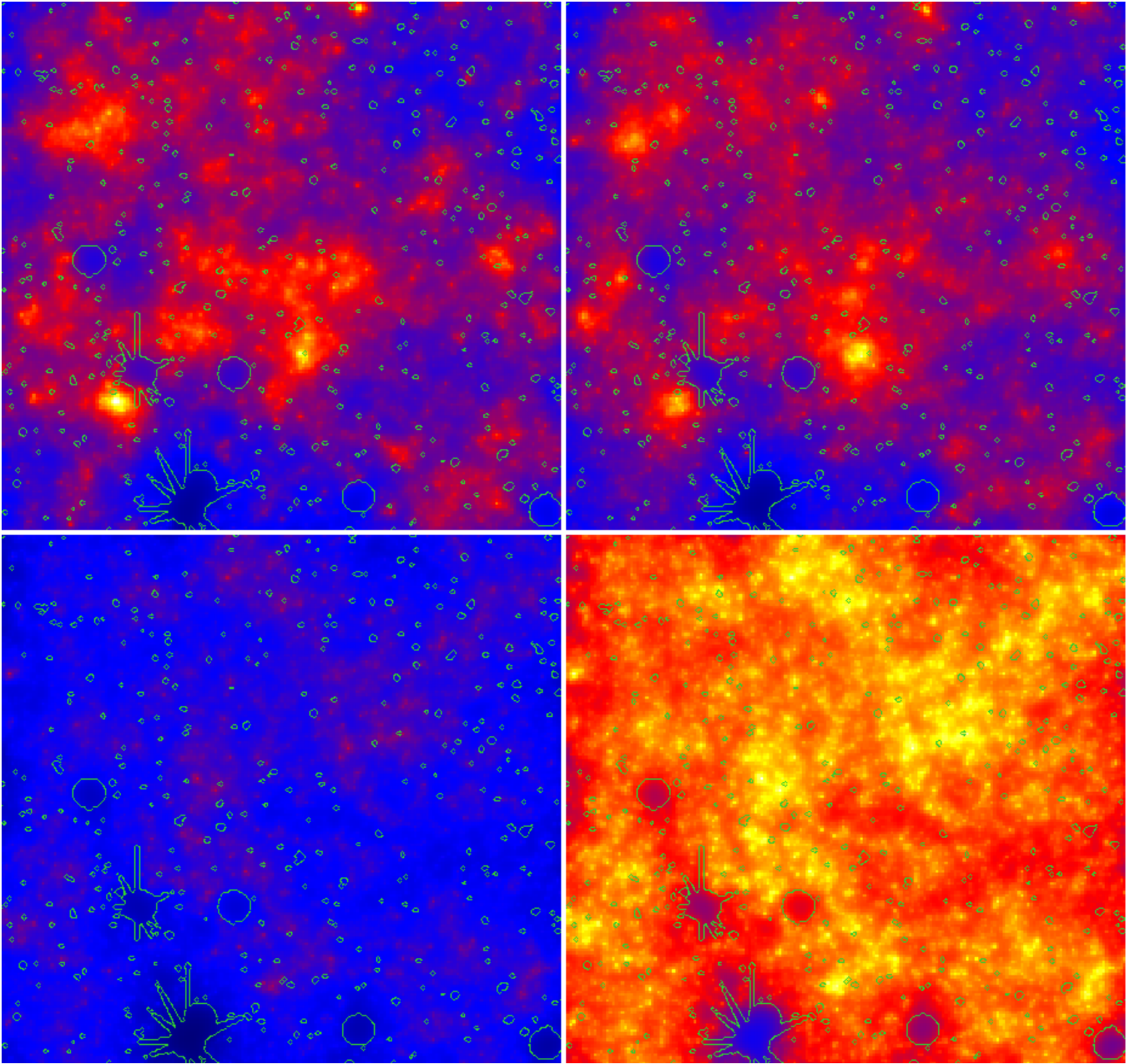

  \centering
  \includegraphics[width=1\textwidth]{{{./fig/test_mocks_179.0_1.5_bis_withmasks}}}
  \caption{Map of the amplitude $A$ of 1 deg$^2$ per side at redshift
    $z=0.35$ for: the original KiDS data (top left), the mock
    catalogue with both field and cluster galaxies (top right), the
    mock catalogue with the field galaxies only and with the same
    color scale of the previous panels (bottom left) and with the
    color scale stretched to better show the details (bottom right).
    The green contour lines outline the areas masking the artefacts
    caused by bright stars and image defects in general.}
  \label{fig:mocks_amplitude}
\end{figure*}

\section{Assessing the quality of the detections}\label{sec:accessing}

In this section, we describe a method to produce realistic mock
catalogues, constructed from the real data themselves, that we use to
estimate the uncertainties of the quantities characterizing the
detections as well as the purity and completeness of the entire
sample.

\begin{figure*}
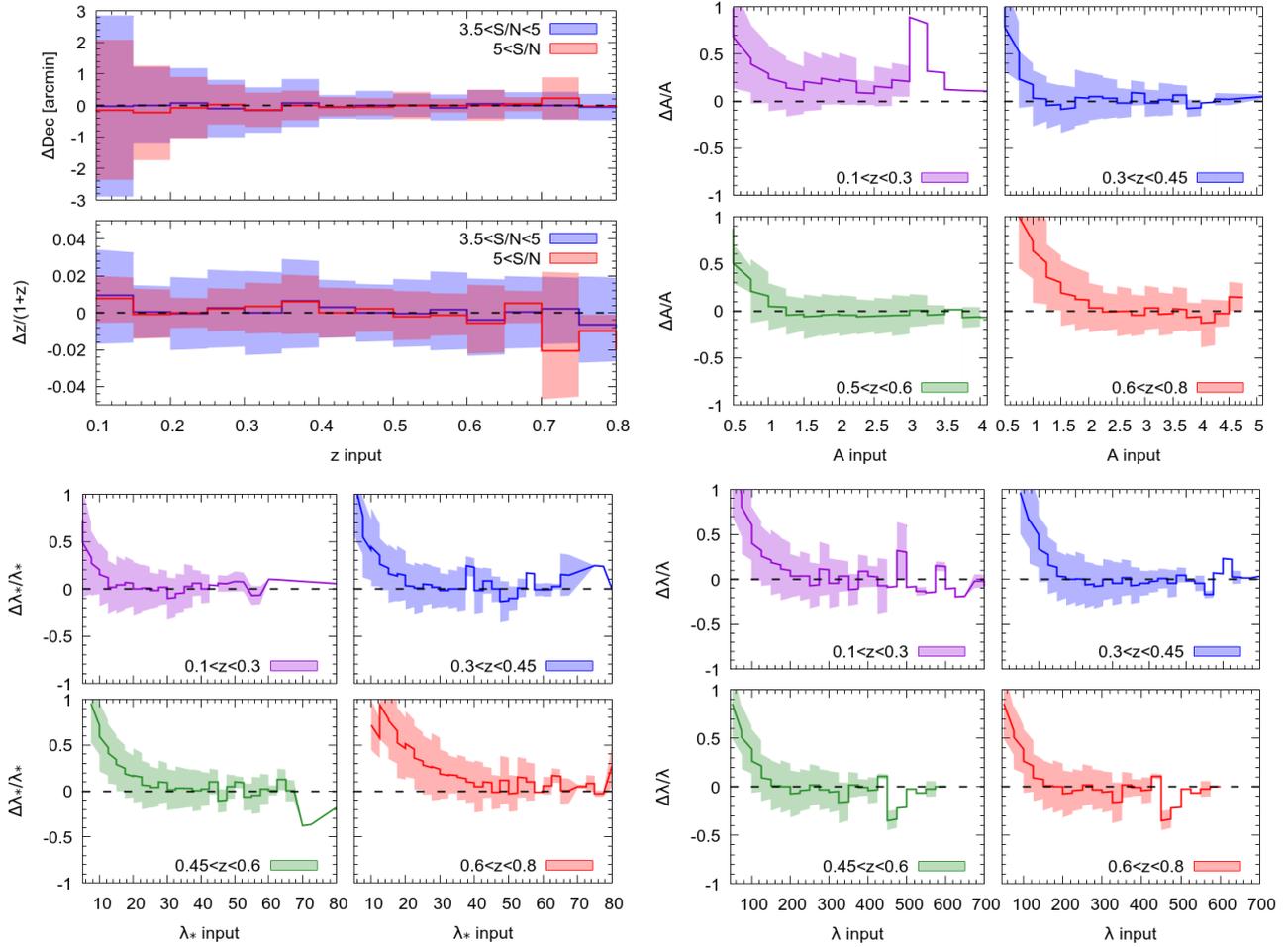

  \centering
  \includegraphics[width=0.48\textwidth]{{{./fig/all_match_dy-z_errors}}}
  \includegraphics[width=0.48\textwidth]{{{./fig/all_match_amplitude_error_minsn3.5_all}}}
  \includegraphics[width=0.48\textwidth]{{{./fig/all_match_richstar_error_minsn3.5_all}}}
  \includegraphics[width=0.48\textwidth]{{{./fig/all_match_rich_error_minsn3.5_all}}}
  \caption{Top left panels: expected statistical $1\sigma$ error
    (shaded areas) on the redshift and angular position (declination)
    as a function of redshift for two different signal-to-noise ratio
    intervals. Relative error of the amplitude $A$ (top right),
    intrinsic richness $\lambda_*$ (bottom left) and apparent richness
    $\lambda$ (bottom right) in four different intervals of redshift.}
  \label{fig:uncertainties}
\end{figure*}

\subsection{Mock simulations}\label{sec:mocks}

We base our mocks on the original KiDS-DR3 data rather then on
numerical simulations to minimize the number of assumptions and to
account for all expected and unexpected properties of the survey
across the sky, such as photometric and phot-z uncertainties,
absorption, masks, variation in depth as well as the clustering of
galaxies etc. The central idea of the mocks rely on a Monte Carlo
extraction of the galaxies based on their probabilistic association to
the entries in our cluster sample, Equation~(\ref{eq:probability}),
and to the field, Equation~(\ref{eq:prob_field}). A scheme with all
steps involved in the mock generation is shown in
Figure~(\ref{fig:scatch-mock}).

In the following, we describe the procedure to create the mock field
population, Section~(\ref{sec:mockField}), and the mock clusters,
Section~(\ref{sec:mock-clusters}). We will discuss several aspects
related to the methodology in Section~(\ref{sec:mock-further}).

\subsubsection{Mock field galaxies}\label{sec:mockField}

The field galaxies are extracted from the KiDS data catalogue via a
Monte Carlo sampling based on the probability, $P_{f,i}$.  In detail,
for each galaxy we extract a uniform random number $r_i$ between 0 and
1, and assign the galaxy to the field if $r_i<P_{f,i}$. For example, a
galaxy with $P_{f}=0.32$ has the $32\%$ chance to be extracted and
associated to the field. All observed properties of these galaxies are
preserved except for their position in the sky which is slightly
perturbed by introducing a random angular displacement. The maximum
random displacement is a free parameter which we set to $r_{rnd}\le1$
Mpc/h. This scale is large enough to dump the presence of clusters
which might have not been detected by the algorithm but is small
enough to preserve the correlation of the Large Scale Structures
(LSS).

\subsubsection{Mock galaxy clusters}\label{sec:mock-clusters}

To generate the mock clusters we started by defining bins of apparent
richness $\lambda$ and redshift $z$ in which to collect all galaxies
associated to clusters. All galaxies with $P_i(j)>0$ have been
considered and those with more than one cluster association have been
attributed to more than one bin accordingly. In this way, each bin
contains all galaxies potentially belonging to clusters with the
richness and redshift defining the bin itself. Each mock cluster is
then generated by randomly extracting galaxies out of the
corresponding bin via a Monte Carlo sampling based on their cluster
membership probability (see Equation~\ref{eq:probability}) and
accounting for the presence of the masked areas in the actual survey
The number of members for each cluster is given by $\lambda$ which is
in fact the number of visible galaxies for that cluster. In short, the
resulting mock cluster is a random realization based on the overall
statistical properties of all original detections with similar
$\lambda$ and $z$.

The mock clusters are then injected into the field mantaining the
angular position, apparent richness and redshift of all cluster
detections with $S/N \ge 3$ found in the original catalogue.  In this
way, we avoid any assumption and rely solely on the statistics of the
data in terms of the correlation of clusters with the LSS and of
clusters with clusters, as well as blending, missing data, non uniform
absorption across the survey, photometric and photo-$z$ uncertainties.
Clearly, clusters which are far and/or small enough to have $S/N<3$
are not generated in our mocks but this does not have a substantial
impact on the results because the final catalogue is limited to the
detections with $S/N \ge 3.5$. In fact, even if objects with lower
$S/N$ would be generated in the mocks, their probability to exceed the
$S/N=3.5$ threshold when measured in the mocks would be very small so
that their contribution to the final sample would anyway be
negligible.

In total, the simulations contains $9018$ mock-clusters over $200$
fields covering a total area of $189$ deg$^2$.

\subsubsection{Further considerations}\label{sec:mock-further}

This mock generation has the advantage of fully preserving all
statistical properties of the original data catalogue by
construction. The overall process boils down to a data bootstrapping
based on the probabilities $P_i(j)$ and $P_{i,f}$ which by
construction sum up to unity. In other words, the mock catalogue is a
random realization based on the original data and only the labelling
of the galaxies (cluster members or field galaxies) mildly depends on
the assumed initial model.

The only assumption within this procedure is hidden in the membership
probabilistic association which directly depends on the cluster model
used to define the filter, see Equation~(\ref{eq:probability}). In
spite of that, such dependency of the mock clusters on the assumed
model is softened by the fact that the mock generation goes through
the original catalogue once more after the detection process has been
completed. This reiteration through the data helps to recover the
radial density distribution and luminosity function of the actual
clusters because the galaxies are used with their real magnitude and
spatial distribution thus mitigating the model assumptions. For
example, let us consider an extreme case in which we assume the
density radial profile of clusters to be flat. In this scenario, the
function describing the membership probabilistic association of
galaxies to clusters has no dependency on the galaxies position.
Consequently, galaxies with different distances from the cluster
center are equally likely to be extracted during the Monte Carlo
sampling but the pool of galaxies out of which we extract the members
has a population which follows the actual density of galaxies, for the
simple reason that these are the actual galaxies in the
data. Therefore, there are more galaxies close to the center to be
sampled then in the outskirts and hence the mock clusters will have a
radial density profile closer to the data than to the initial
assumption. The mock clusters are not a mere representation of the
model.

We show in the top left panel of Figure~(\ref{fig:mocks_amplitude})
the amplitude map for a slice at redshift $z=0.35$ resulting from the
analysis of about 1 deg$^2$ per side of the real KiDS data (top left
panel). The green contour lines outline the areas masking the
artefacts caused by bright stars and image defects in general. In the
top right panel, we show the same analysis but performed on the
corresponding mock catalogue of galaxies. All main features are
clearly preserved and the small differences between the left and right
maps are due both to the displacement of the galaxy positions, and to
the Monte-Carlo sampling process. In practice, our mock map is a
realization of the population of galaxies, statistically independent
from the original. In the two bottom panels of the same figure, we
plot with two different color scales the contribution to the amplitude
of the mock field galaxies alone. The two top panels and bottom left
one have the same color scale, while in the right bottom panel we
stretched the contrast to better highlight the details such as the LSS
pattern.

\subsection{Uncertainties on the detections properties}\label{sec:uncertainties}

The uncertainties on the properties of the detections (position,
redshift, amplitude, richness, etc...) are evaluated through the
analysis of the mock catalogues by running on them the AMICO code as
done with the real data and comparing the measured values with the
expected ones. The errors (1 $\sigma$ uncertainties) estimated in this
way are listed in the final catalogue of detections. One detection is
assigned to one mock cluster present in the simulations if they lay
within a cylinder of $\Delta r\le 1 \mbox{Mpc/h}$ in radius and of
$\Delta z=0.1$ in length. The detections without a match are
considered as spurious and allow us to derive the purity of the
sample. The results are presented in Figure~(\ref{fig:uncertainties})
and are discussed below.

\begin{figure}
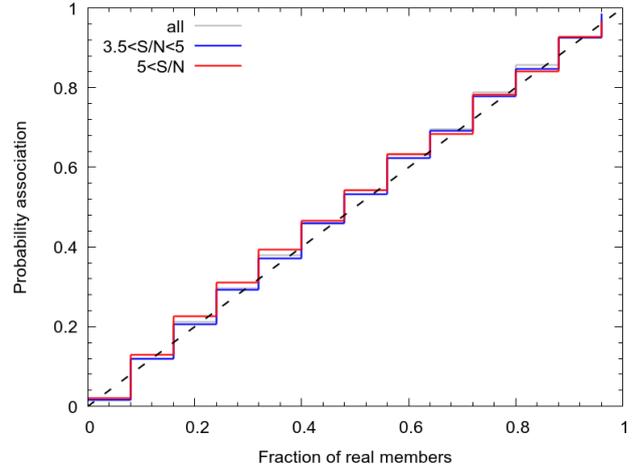

  \centering
  \includegraphics[width=0.48\textwidth]{{{./fig/all_association}}}
  \caption {Membership probabilistic association of the galaxies
    against the fraction of actual members.}
  \label{fig:association}
\end{figure}

{\it Angular position}: the scatter along the declination, $\Delta
Dec$, of the detections is larger for clusters at lower redshift
because of their larger angular extension which does not allow for a
well defined localization of their center. At higher redshift,
$z>0.45$, the angular resolution is dictated by the pixels size we
have chosen for the maps produced by AMICO, that is of $\sim0.3$
arcmin. The scatter along R.A. is completely analogous and because of
this we do not shown it.

\begin{figure*}
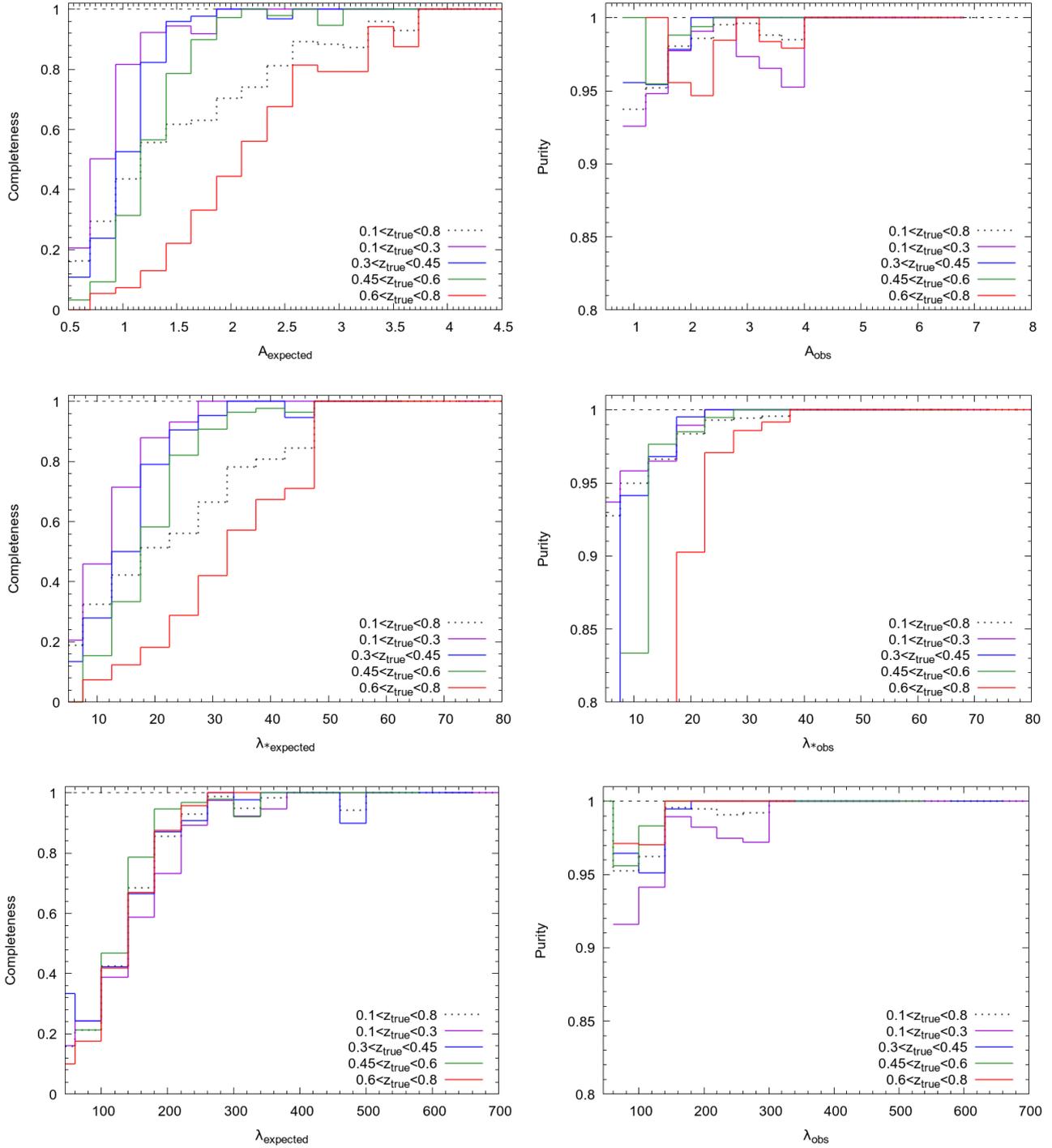

  \centering
  \includegraphics[width=0.48\textwidth]{{{./fig/all_completeness_amplitude_z_min_sn3.5}}}
  \includegraphics[width=0.48\textwidth]{{{./fig/all_purity_amplitude_z_minsn3.5}}}

  \includegraphics[width=0.48\textwidth]{{{./fig/all_completeness_richstar_z_min_sn3.5}}}
  \includegraphics[width=0.48\textwidth]{{{./fig/all_purity_richstar_z_minsn3.5}}}
  
  \includegraphics[width=0.48\textwidth]{{{./fig/all_completeness_rich_z_min_sn3.5}}}
  \includegraphics[width=0.48\textwidth]{{{./fig/all_purity_rich_z_minsn3.5}}}
  \caption{Completeness (left panels) and purity (right panels) for
    four different redshifts intervals as a function of amplitude $A$,
    intrinsic richness $\lambda_*$ and apparent richness $\lambda$
    from top to bottom, respectively. The sample completeness, with
    respect to the amplitude $A$ and intrinsic richness $\lambda_*$,
    changes with redshift. This is not the case when the apparent
    richness $\lambda$ is adopted.
  }
  \label{fig:purity-completenss}
\end{figure*}

\begin{figure*}
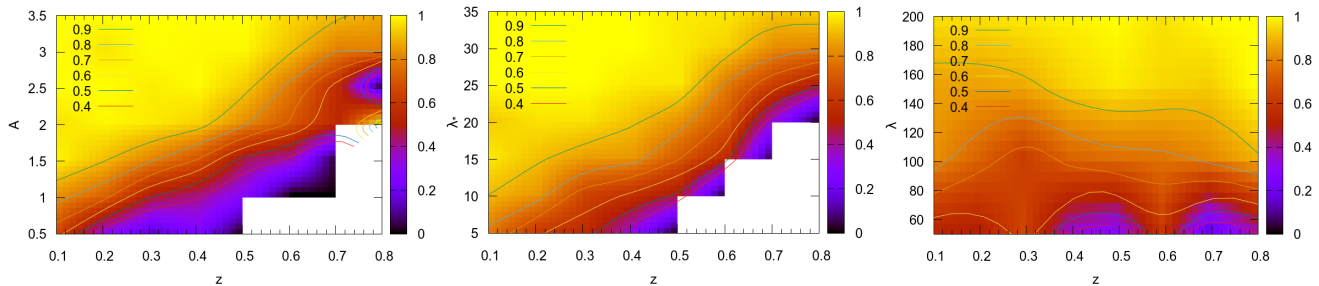

  \centering
  \includegraphics[width=0.32\textwidth]{{{./fig/all_completeness_2D_amplitude}}}
  \includegraphics[width=0.32\textwidth]{{{./fig/all_completeness_2D_lambdastar}}}
  \includegraphics[width=0.32\textwidth]{{{./fig/all_completeness_2D_lambda}}}
  \caption{Left panel: completeness of the cluster catalogue as a
    function of redshift $z$ and amplitude $A$. Center and right
    panel: same as left panel but with intrinsic richness $\lambda_*$
    and apparent richness $\lambda$, respectively, instead of
    amplitude. The iso-contours trace the completeness from 0.4 to
    0.9. Where the completeness is extremely low there are basically
    no detections and therefore no sufficient data (white areas).}
  \label{fig:completenss-2d}
\end{figure*}

{\it Redshift}: the relative scatter in redshift is constant over the
whole redshift interval and it amounts to $\Delta z/(1+z) \sim 0.02$,
which is smaller then the one of the galaxy population $\Delta_{gal}
z/(1+z) \sim 0.044$ \citep{deJong17}. The known bias of the galaxies
photo-$z$ already discussed in \cite{deJong17} surely affects the
redshift estimates of the detections but this does not show up in the
top left panel of Figure~(\ref{fig:uncertainties}). This is because
the reference redshift of the mock clusters has been taken from the
data and is therefore dragged by the very same bias. Nevertheless,
this test serves to estimate the redshift uncertainty in a reliable
way and shows that AMICO returns unbiased results with respect to the
input photo-$z$ catalogue. We decided not to apply any correction to
the redshifts of our sample to leave it as a separate step as more
data against with to calibrate become available with time and the
understanding of the photo-$z$ of the galaxies improves. The final
redshift correction is thus left to the user but, based on the
analysis discussed in Section~(\ref{sec:gama}), we suggest to apply
the following relation to our sample $z_{corrected} = z - 0.02\,(1+z)$
in agreement with \cite{deJong17}. Our approach differs from the one
adopted in redMaPPer where such calibration against external
spectroscopic data is done internally \citep{Rykoff14}.

{\it Amplitude}: the relative scatter of the amplitude $A$ is almost
constant over the whole range of values and in all four redshift
intervals displayed in the top right four panels of
Figure~(\ref{fig:uncertainties}). At $z>0.3$, the amplitude is biased
down to a minimum value, which grows with redshift, below which the
sample becomes incomplete and the Malmquist bias manifests itself as
expected. In contrast, the amplitude of the lower redshift bin is
biased high. We verified that this is not due to the masks which are
more relevant for the lower redshift clusters because of their larger
angular extension. The origin of this amplitude bias is not completely
understood;

{\it Richness}: similar considerations hold for the intrinsic richness
$\lambda_*$ and apparent richness $\lambda$ except for the fact that
these quantities are unbiased at all redshifts.

{\it Membership probabilistic association}: in Figure~(\ref{fig:association}) we
show the fraction of real members as a function of the measured
probabilistic association of the galaxies to clusters (see
Equation~\ref{eq:probability}). Two different intervals of
signal-to-noise ratio, that are $3.5<S/N<5$ and $S/N>5$, are
shown. The correlation well satisfies the identity proving that the
association of galaxies to clusters is reliably estimated.

\subsection{Purity and completeness}\label{sec:completeness}

The completeness and the purity of the final sample are shown in
Figure~(\ref{fig:purity-completenss}). The purity is a measure of the
contamination level of the cluster sample and is defined as the
fraction of detection successfully matched with the clusters in the
simulations over the total number of detections. This is shown in the
right panels of Figure~(\ref{fig:purity-completenss}). The sample
purity is extremely high. The completeness, shown in the left panels,
is defined as the fraction of detections with the fraction of mock
clusters with a given amplitude or richness which have been detected
correctly identified as clusters over the total number of mock
clusters in the simulations. The minimum amplitude $A$ and intrinsic
richness $\lambda_*$ for which the sample is complete grows with
redshift. This is because these two quantities are direct estimates of
the clusters mass and, clearly, the larger is the redshift, the larger
is the minimum mass for a cluster to be detected above a fixed minimum
signal-to-noise ratio, that is $S/N>3.5$ in our case. This is not the
case for the apparent richness, $\lambda$, because it quantifies the
number of galaxies visible in a cluster that is mostly determined by
the depth of the data. This missing or weak redshift dependence makes
$\lambda$ a better probe for cosmological studies. In
Figure~(\ref{fig:completenss-2d}), we show the completeness as a
function of redshift for different levels of amplitude $A$, intrinsic
richness $\lambda_*$, and apparent richness $\lambda$. Note that the
completeness is measured not by setting a fixed mass threshold for all
redshifts but instead it refers to a richness limit which on average
grows with redshift. This definition is a consequence on how the mocks
have been constructed. The population of clusters below such threshold
is observationally un-accessible and could only be evaluated by
assuming, for example, a mass function and a model relating dark
matter haloes to visible galaxies, but such a study is not of our
interest because fully model-dependent. What this method is aiming at
is a model-independent selection function, related to completeness and
purity, based on observables only that can then be used to
investigate, for instance, the cosmological model or the star
formation history. In fact, the mass proxies discussed in this paper
can be related to actual masses for a direct comparison with
theoretical models thanks to the scaling relations based on
weak-lensing mass measurements derived in Bellagamba et al. submitted.

\begin{table*}
  \caption{Entries listed in the catalogue of galaxy clusters.}
  \centering
  \begin{tabular}{ll}
    \hline
    NAME        & unique identification name: AMICO-KIDS3-\#\\
    ID          & unique identification number\\
    FIELD       & identification number ot the tile in which the detection has been found\\
    XPIX, YPIX, ZPIX & the indexes of the pixel of the amplitude map in which the detection falls\\
    XSKY, YSKY, ZSKY & sky coordinates R.A., Dec. and the redshift corresponding to XPIX, YPIX and ZPIX\\
    A   & amplitude, $A$, as defined in Equation~(\ref{eq:amplitude_new})\\
    LAMBDA      & apparent richness, $\lambda$, as defined in Equation~(\ref{eq:apprichness})\\
    LAMBDASTAR  & intrinsic richness, $\lambda_*$, as defined in Equation~(\ref{eq:starrichness})\\
    XPIX\_ERR, YPIX\_ERR, ZPIX\_ERR & 1 $\sigma$ error of the position in pixel units \\
    XSKY\_ERR, YSKY\_ERR, ZSKY\_ERR & 1 $\sigma$ error of the position in R.A., Dec. and the redshift\\
    A\_ERR      & 1 $\sigma$ error of the amplitude defined in Equation~(\ref{eq:amplitudesigma})\\
    LAMBDA\_ERR & 1 $\sigma$ error of $\lambda$ based on the mock catalogues\\
    LAMBDASTAR\_ERR & 1 $\sigma$ error of $\lambda_*$ based on the mock catalogues\\
    SN          & signal-to-noise ratio based on the amplitude, AMPLITUDE, and its r.m.s, A\_ERR\\
    LIKELIHOOD  & likelihood derived in Equation~(\ref{eq:likelihood})\\
    MASKFRAC    & fraction of the detection which is masked\\
    ID\_LITERATURE & identification number for those clusters already listed in the literature\\
    \hline
  \end{tabular}
  \label{tab:entries_list}
\end{table*}

\section{Conclusions}\label{sec:conclusions}

We detected galaxy clusters in the KiDS-DR3 data with the AMICO
code. In the analysis, we avoided those regions of the sky affected by
the presence of artefacts produced, for example, by bright stars and
image artefacts, thus covering an effective area of $377$
deg$^2$. With respect to our previous study of the KiDS-DR2
\citep{Radovich17}, the work presented here takes advantage of the
improvements with respect to the standard matched filter method
introduced with the AMICO detection algorithm \citep{Bellagamba18},
such as the cleaning procedure, a probability redshift distribution of
the filter which now depends on the individual magnitude of the
galaxies and a more robust approach to deal with possible biases in
the galaxies photo-$z$s. We detected $7988$ galaxy clusters over a
redshift range of $0.1<z<0.8$ with a minimum signal-to-noise ratio of
$S/N=3.5$. The catalog lists for each detection its unique
identification number, sky position, redshift, amplitude $A$,
intrinsic richness $\lambda_*$, apparent richness $\lambda$,
signa-to-noise ratio, likelihood $mathcal{L}$, masked fraction, full
probability redshift distribution and its name in the literature if
present. In the process we also derived the probabilistic association
of galaxies to each cluster, a useful information to study galaxy
formation or help in the removal of the foreground for weak-lensing
studies, just to mention some applications.

We compared our sample to public and private catalogues of galaxy
clusters overlapping our fields, in particular:
(1) we matched the cluster candidates detected by the redMaPPer
algorithm on the Sloan Digital Sky Survey Data Release 8 (SDSS-DR8);
(2) we detected all 19 Planck SZ-selected clusters present in our sky
area and provide, for the first time, a redshift estimate for 10 of
these objects;
(3) we used the X-rays derived masses listed in the MCXC sample of
clusters to test our mass proxies. Even if the set of common objects
is not large enough to definitive conclusive results, the clear
correlation we find is extremely encouraging;
(4) finally, we used the GAMA-I galaxy group catalogue (G$^3$C) to
verify our redshift estimate. This allowed us to confirm the already
known photo-$z$ bias affecting the KiDS data and derived the required
calibration for its correction.

Finally, we proposed a new methodology based on the galaxy membership
probabilistic association provided by AMICO to create realistic mock
catalogues and use them to evaluate the uncertainties of all the
properties of the detections, such as their angular position in the
sky, redshift, and mass proxies. Most importantly, we use this method
to derive the selection function of the sample in a fully model
independent way. As it turned out, the sample has an extremely high
purity, approaching $90$\% over the whole redshift range. This method
provides the first step towards the measure of cosmological parameters
through the use of photometrically selected galaxy clusters.

The catalogue of clusters and of the cluster members will be made
publicly available but they can already be requested after submitting
a proposal.

\appendix

\section{Some example of detections}\label{sec:stamps_examples}

Here we show a sample of 12 detections located at four different
redshifts ($z=0.2$, $0.3$, $0.5$ and $0.7$, rows from top to bottom),
and three different intrinsic richnesses ($\lambda_*\approx 100$, $50$
and $5-10$, column from left to right). All cut-outs are of $400$
arcsec in size.

\begin{figure*}
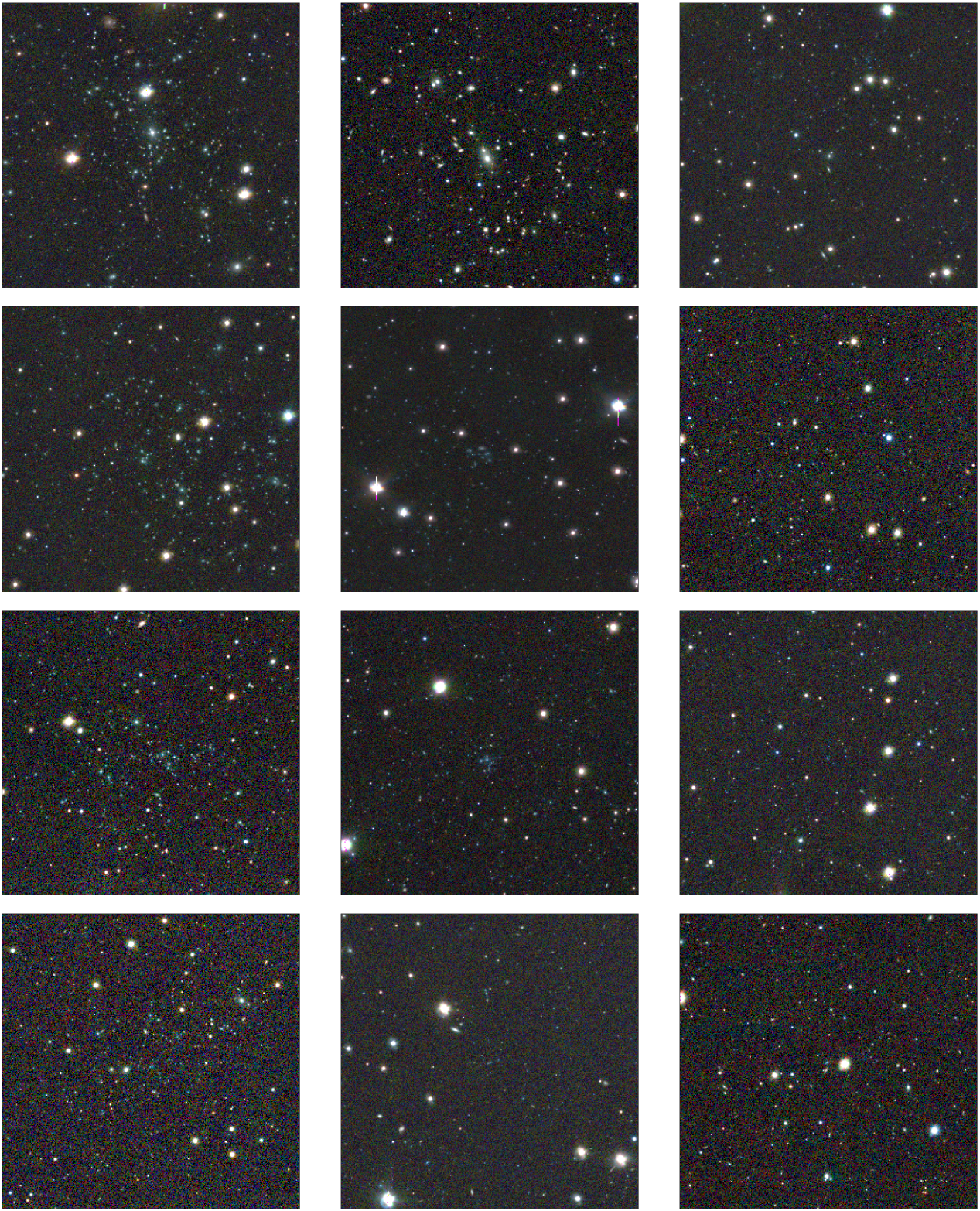

  \centering
  \includegraphics[width=0.32\textwidth]{{{./fig/stamp_4141_notics_contrast}}}
  \includegraphics[width=0.32\textwidth]{{{./fig/stamp_2342_notics}}}
  \includegraphics[width=0.32\textwidth]{{{./fig/stamp_471_notics_contrast}}}
    
  \vspace{-0.5cm}
  \includegraphics[width=0.32\textwidth]{{{./fig/stamp_1245_notics_contrast}}}
  \includegraphics[width=0.32\textwidth]{{{./fig/stamp_481_notics_contrast}}}
  \includegraphics[width=0.32\textwidth]{{{./fig/stamp_7280_notics}}}

  \vspace{-0.5cm}
  \includegraphics[width=0.32\textwidth]{{{./fig/stamp_2593_notics}}}
  \includegraphics[width=0.32\textwidth]{{{./fig/stamp_6686_notics_contrast}}}
  \includegraphics[width=0.32\textwidth]{{{./fig/stamp_6355_notics_contrast}}}

  \vspace{-0.5cm}  
  \includegraphics[width=0.32\textwidth]{{{./fig/stamp_2049_notics}}}
  \includegraphics[width=0.32\textwidth]{{{./fig/stamp_3934_notics_contrast}}}
  \includegraphics[width=0.32\textwidth]{{{./fig/stamp_7198_notics}}}

  \caption{A sample of 12 detections located at redshifts $z=0.2$,
    $0.3$, $0.5$ and $0.7$ (rows from top to bottom), and with an
    intrinsic richnesses of $\lambda_*\approx 100$, $50$ and $5-10$
    (columns from left to right). All postage-stamps have a size of
    $400$ arcsec.}
  \label{fig:examples}
\end{figure*}

\subsection*{Acknowledgements}

This work was supported by the Collaborative Research Center TR33 ’The
Dark Universe’. FB, MRo and LM thank the support from the grants ASI
n.I/023/12/0 “Attivit\'a relative alla fase B2/C per la missione
Euclid” and PRIN MIUR 2015 “Cosmology and Fundamental Physics:
Illuminating the Dark Universe with Euclid”. MS acknowledges financial
support from the contracts ASI-INAF I/009/10/0, NARO15 ASI-INAF
I/037/12/0, ASI 2015-046-R.0, and ASI-INAF n.2017-14-H.0.


\end{document}